\documentclass[conference]{IEEEtran}
\IEEEoverridecommandlockouts
\usepackage{amsmath, amsfonts, amssymb, amsthm}
\usepackage{algorithm}
\usepackage{algpseudocode}
\usepackage{graphicx}
\usepackage{epstopdf}
\usepackage{textcomp}
\usepackage{xcolor}
\usepackage[nolist]{acronym}
\usepackage{bm}
\usepackage{afterpage}
\usepackage{subcaption}
\usepackage[utf8]{inputenc}
\usepackage{cite}
\usepackage{geometry}
\allowdisplaybreaks
\title{Near-Field THz Bending Beamforming: A Convex Optimization Perspective}

\author{
	\IEEEauthorblockN{Aoran Liu\textsuperscript{*}, Weidong Mei\textsuperscript{*}, Peilan Wang\textsuperscript{*}, Dong Wang\textsuperscript{*}, Ya Fei Wu\textsuperscript{†}, Zhi Chen\textsuperscript{*}, Boyu Ning\textsuperscript{*}}
	\IEEEauthorblockA{\textsuperscript{*} National Key Laboratory of Wireless Communications,} 
	\IEEEauthorblockA{\textsuperscript{†} EHF Key Laboratory of Fundamental Science, School of Electronic Science and Engineering, \\
		University of Electronic Science and Technology of China, Chengdu 611731, China}
	\IEEEauthorblockA{Emails: liuaoran@std.uestc.edu.cn; wmei@uestc.edu.cn; peilan\_wangle@uestc.edu.cn;\\ DongwangUESTC@outlook.com; wuyafei@uestc.edu.cn; chenzhi@uestc.edu.cn; boydning@outlook.com}
}
\begin{document}

\maketitle

\begin{acronym}
	\acro{THz}{Terahertz}
	\acro{BS}{base stations}
	\acro{6G}{sixth-generation}
	\acro{ABF}{analog beamforming}
	\acro{DBF}{digital beamforming}
	\acro{SCA}{successive convex approximation}
	\acro{SVD}{singular value decomposition}
	\acro{SDR}{semidefinite relaxation}
	\acro{MIMO}{multiple-input multiple-output}
	\acro{LoS}{line-of-sight}
	\acro{ELAA}{extremely large antenna arrays}
	\acro{EM}{electromagnetic}
	\acro{NLoS}{non-line-of-sight}
	\acro{SDP}{semi-definite programming}
	\acro{TM}{tangent method}
	\acro{AP}{access point}
\end{acronym}
\begin{abstract}
\par \ac{THz} communication systems suffer severe blockage issues, which may significantly degrade the communication coverage and quality. Bending beams, capable of adjusting their propagation direction to bypass obstacles, have recently emerged as a promising solution to resolve this issue by engineering the propagation trajectory of the beam. However, traditional bending beam generation methods rely heavily on the specific geometric properties of the propagation trajectory and can only achieve sub-optimal performance. In this paper, we propose a new and general bending beamforming method by adopting the convex optimization techniques. In particular, we formulate the bending beamforming design as a max-min optimization problem, aiming to optimize the analog or digital transmit beamforming vector to maximize the minimum received signal power among all positions along the bending beam trajectory. However, the resulting problem is non-convex and difficult to be solved optimally. To tackle this difficulty, we apply the \ac{SCA} technique to obtain a high-quality suboptimal solution. Numerical results show that our proposed bending beamforming method outperforms the traditional method and shows robustness to the obstacle in the environment.
\end{abstract}
	\section{Introduction}
	\par \acf{THz} communication has emerged as a key enabler for next-generation wireless systems that demand ultra-high data rates\cite{boyu_thz}. However, due to the substantial path loss at \ac{THz} frequencies, the use of large or even \ac{ELAA} becomes crucial for generating high-gain beams to ensure wireless coverage \cite{pencil_like_beam}. The expanded aperture of \ac{ELAA} extends the near-field (Fresnel) region to a large distance—potentially exceeding 100 meters at 300 GHz \cite{300GHz_100m}. While this extended near-field provides a unique beamfocusing capability of concentrating \ac{EM} energy on sub-wavelength focal spots, it also introduces significant vulnerability to blockages. Even centimeter-scale obstructions in the \ac{LoS} direction can catastrophically disrupt the wireless communication performance. Although several mitigation strategies, such as ultra-dense networks (UDNs) \cite{Ultra-Dense_Network1,Ultra-Dense_Network2} and reconfigurable intelligent surface/intelligent reflecting surface (RIS/IRS) \cite{Intelligent_Reflecting_Surface,weidong_ris}, have been proposed in the literature, they all face various challenges in practice, including the high-cost hardware deployment, difficulty in interference management, high coordination overhead, among others.
	\begin{figure}[htbp]
		\centering
		\includegraphics[width=0.45\textwidth]{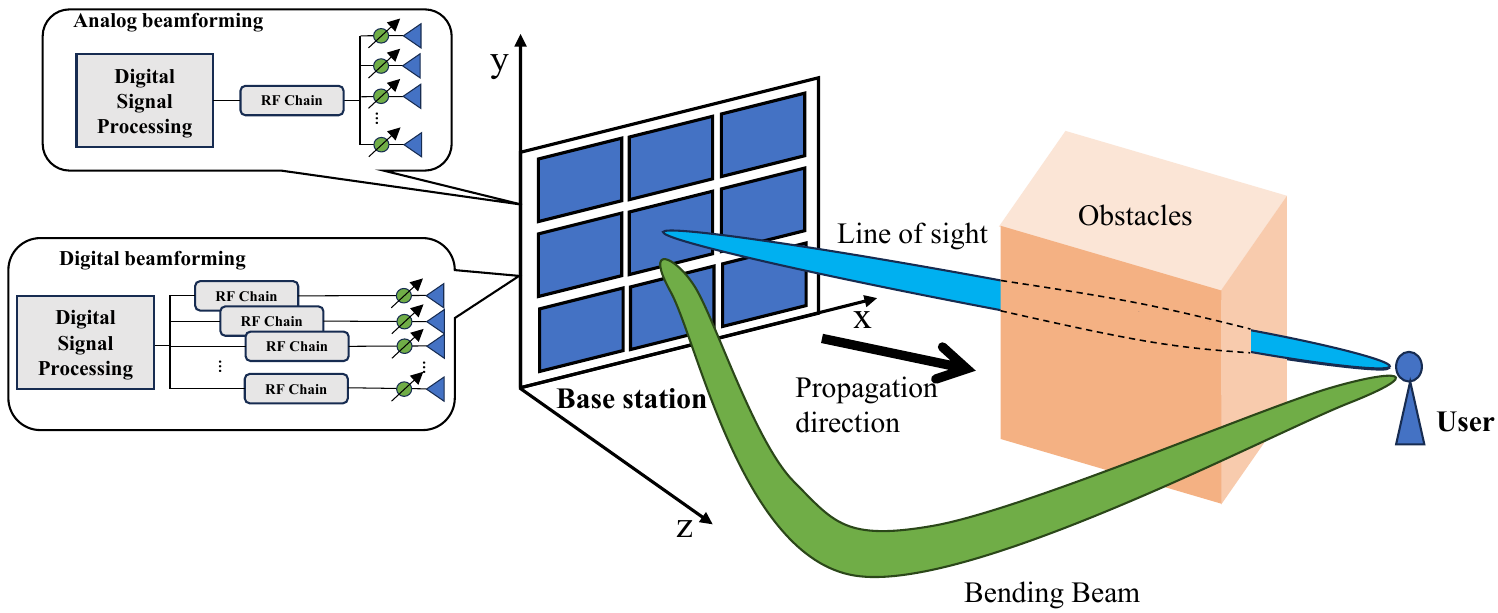} 
		\caption{Bending beam generation via analog/digital beamforming.}
		\label{fig:bending beam} % 图片的标签，用于交叉引用
		\vspace{-12pt}
	\end{figure}
		\par Fortunately, recent advances in wavefront engineering have demonstrated the potential of non-diffracting beams for achieving reliable wireless communications even with obstacles in the \ac{LoS} direction, by exploiting their inherent ``self-healing" properties. For example, one typical non-diffracting beam is referred to as bending beam, which is able to focus the beam energy on a prescribed curved trajectory in the non-\ac{LoS} direction, as illustrated in Fig. \ref{fig:bending beam}. By this means, a \ac{BS}/\ac{AP} can achieve consistently reliable communications with dynamically blocked users, thus significantly enhancing the \ac{THz} signal coverage, especially in the complex environment with dense obstacles. The fundamental physics of such bending beams can be characterized through the stationary solutions to the celebrated Helmholtz equation, which governs the spatial evolution of \ac{EM} wavefront\cite{Helmholtz_equ,wavefont,wavefonthopping}. A pioneering work \cite{nature_comunication} has demonstrated \ac{THz} bending beam using an \ac{ELAA} system. However, this method can only generate a specific type of trajectory for the bending beam, i.e. Airy beam, by following the Airy function, which may limit its applications in practice. To enable more flexible bending beamforming, a recent work \cite{6G_bending_beam} proposed an alternative method by assuming a continuous antenna aperture and determining the phase profile at any point based on its tangent line to the desired trajectory. However, this method may still achieve suboptimal performance in general, due to its reliance on the geometric properties of the desired trajectory and other limitations, as will be detailed in this paper later.
		\par To address the above challenges, we propose a new and general bending beamforming method by adopting the convex optimization techniques. Unlike the physics-driven approaches as in \cite{nature_comunication} and \cite{6G_bending_beam}, we formulate the bending beamforming design as an optimization problem. Specifically, we aim to maximize the minimum received signal power among all positions along the desired propagation trajectory by optimizing the digital or analog beamformer at the \ac{BS}. However, the resulting problem is non-convex, making it difficult to be solved optimally. To address this difficulty, we employ the \acf{SCA} technique to obtain a high-quality suboptimal solution by solving a series of approximate convex optimization problems iteratively. Numerical results demonstrate that our proposed bending beamforming method outperforms the traditional method and shows robustness to the obstacles in the environment.
			\par \textit{Notations}: $\mathbb{R}$ and $\mathbb{C}$ represent the sets of real and complex numbers, respectively. For a complex number $a$, $\operatorname{Re}\{a\}$ and $|a|$ denote its real part and modulus, respectively. For a complex-valued vector $\mathbf{x}$, $\mathbf{x}^\text{H}$, $\|\mathbf{x}\|_2$, and $x(n)$ denote its transpose, conjugate transpose, $\ell_2$-norm, and the $n$-th element, respectively. For a matrix $\bm{A}$, $\operatorname{Tr}(\bm{A})$ represents its trace. $f'(x)$ denotes the first-order derivative of $f(x)$.
	\section{System Model}
	\par As shown in Fig. \ref{fig:bending beam}, we consider the downlink transmission from a multi-antenna \ac{BS} to a single-antenna user. The BS is assumed to be equipped with a uniform linear array (ULA) with its length assumed to be $L$ meter (m). Due to the blockage in the environment, the \ac{BS} aims to generate a bending beam to bypass the blockage and serve the user accordingly. For ease of exposition, we establish a three-dimensional (3D) Cartesian coordinate system, as depicted in Fig. \ref{fig:bending beam}. In this paper, we consider two beamforming schemes adopted by the \ac{BS}, i.e., \ac{ABF} and \ac{DBF}. In particular, in the case of \ac{ABF}, the transmit beamforming vector can be expressed as
	\begin{equation}
		\bm{\omega}_a = \frac{1}{\sqrt{N}}[e^{j\phi_1},e^{j\phi_2},...,e^{j\phi_n},...,e^{j\phi_N}]^\text{T} \in \mathbb{C}^{N\times1}\text{,}
	\end{equation}
	where $\phi_n$ denotes the phase shift at the $n$-th antenna. While in the case of \ac{DBF}, the phase shift and amplitude of each transmit weight can be jointly adjusted. Accordingly, the transmit beamforming vector can be expressed as
		\begin{equation}
		\bm{\omega}_d = [\alpha_1e^{j\phi_1},\alpha_2e^{j\phi_2},...,\alpha_n e^{j\phi_n},...,\alpha_N e^{j\phi_N}]^\text{T} \in \mathbb{C}^{N\times1}\text{,}
	\end{equation}
	where $\alpha_n$ denotes the amplitude at the $n$-th antenna.
	\par To generate the bending beam, we adopt the near-field channel model under the spherical-wave propagation condition. Specifically, in the absence of the blockage, the channel between the $n$-th antenna and any target location in the three-dimensional (3D) space (e.g., $(x,y,z)$) is determined by the free-space propagation, i.e.,
	\begin{equation}
	h_n(x,y,z) = \frac{\lambda}{4\pi d_n(x,y,z)} e^{\frac{j2\pi d_n(x,y,z)}{\lambda}}\text{,}
	\end{equation}
	 where $\lambda$ represents the wavelength of the transmitted signal,  $d_n(x,y,z)$ represents the distance between the target position and the $n$-th antenna. As a result, the channel between of all antennas and the target position can be expressed as
	 \begin{equation}
	 \bm{h}(x,y,z) = [h_1(x,y,z),h_2(x,y,z),...,h_N(x,y,z)]^\text{T}\in \mathbb{C}^{N\times1}\text{,}
	 \end{equation}
	 and the received signal power at any position $(x,y,z)$ is given by
	 \begin{equation}
	 	p(x,y,z) = |\bm{\omega}^\text{H}\bm{h}(x,y,z)|^2\text{,}
	 \end{equation}
	 where $\bm{\omega} \in\left\{\bm{\omega}_{a}, \bm{\omega}_{d}\right\}$. Notably, in the case without the blockage in Fig. \ref{fig:bending beam}, to maximize the received signal power at the position $(x,y,z)$, the optimal beamforming can be determined as an \ac{ABF} vector, i.e., $\bm{h}(x,y,z)/\lVert{\bm{h}}(x,y,z) \rVert$. However, this may cause significant signal attenuation if the blockage is present in the \ac{LoS} direction, especially for ultra-high frequencies (e.g., \ac{THz} \cite{boyu_thz}) with severe penetration losses. Next, we first present the conventional bending beamforming method, followed by the proposed general bending beamforming method relying on convex optimization.
\section{Traditional Bending Beam Generation}
\label{sec:tgb}
\par From the perspective of \ac{EM} theory, bending beam can be characterized by Maxwell’s equations, which for free-space propagation reduce to the Helmholtz equation, i.e.,
\begin{equation}
	\label{equ:Helmholtz_equation}
	(\nabla^2+k^2)\bm{E}(\bm{r})={\bf 0},
\end{equation}
where $\nabla^2$ is the Laplacian operator, $k=2\pi/\lambda$ is the wavenumber, and $\bm{E}(\bm{r})$ is the vector of electric field at any position $\bm{r}=(x,y,z)$ that can be decomposed into three directions along $x$-, $y$-, and $z$-axes, denoted as $E_x$, $E_y$, and $E_z$, respectively. For simplicity, consider a one-dimensional (1D) case, e.g., the electric field is invariant along $y$-axis and propagates along $z$-axis. In this case, $\bm{E}(\bm{r})$ degrades into a scalar ${E}(x, z)e^{jkz}$, and (\ref{equ:Helmholtz_equation}) can be simplified as
\begin{equation}
	\label{equ:helmo}
	j\frac{\partial {E}(x, z)}{\partial z}+\frac{1}{2k}\frac{\partial^2{E}(x, z)}{\partial x^2}=0,
\end{equation}
where $j$ denotes the imaginary unit with $j^2=-1$, and ${E}(x,z)$ denotes the magnitude of the electric field that removes the component in $y$-axis. The solution to (\ref{equ:helmo}) can be expressed as an Airy solution (thus naming its generated bending beam as Airy beam), i.e.,
	\begin{equation}
		\label{equ:airy}
		E(x,z) = E_0{\cal A}\left[(4\beta k^2)^{1/3}\left(x-\beta z^{2}+j \frac{\alpha}{k} z\right)\right] e^{j \phi(x, z)},
	\end{equation}
where  $\phi(x, z)=\left\lceil 2 \beta k\left(x-(2 / 3) \beta z^{2}\right)+\alpha^{2} / 2 k\right\rceil-j \alpha\left(x-2 \beta z^{2}\right)$, $E_0$ is a complex constant, ${\cal A[\cdot]}$ represents Airy function, and $\alpha$ and $\beta$ are real constants. However, the Airy beam in (\ref{equ:airy}) can only follow a parabolic trajectory expressed as $x=\beta z^{2}$ along the beam propagation direction.
\begin{figure}[!t]
	\centering
	\includegraphics[width=0.4\textwidth]{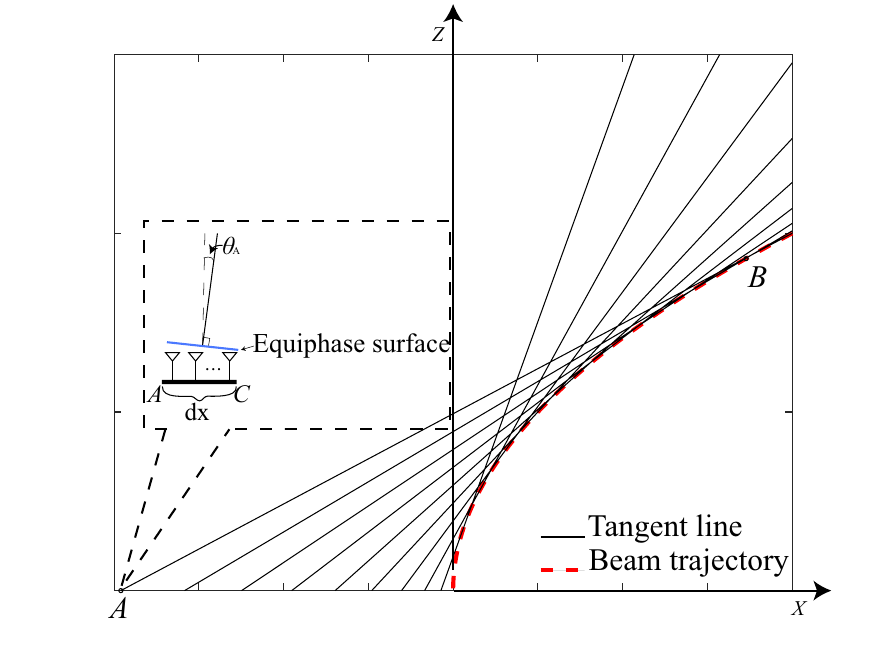} 
	\caption{Illustration of the tangent method.}
	\label{fig:Equivalent_rays} % 图片的标签，用于交叉引用
	\vspace{-8pt}
\end{figure}
\par To generate a bending beam of other shapes, a tangent method was proposed in \cite{6G_bending_beam}. As shown in Fig. \ref{fig:Equivalent_rays}, this method treats the antenna aperture as a continuous array composed of an infinite number of small sub-arrays, each with a length of ${\text d}x$. Then, any point over the beam trajectory can be viewed as being located in the far-field of each small sub-array. To determine the phase shift at any position within the antenna array, we take an arbitrary point $A$ as an example. Specifically, we draw a line through $A$ that is tangent to the beam trajecotry at point $B$, as shown in Fig. \ref{fig:Equivalent_rays}. Then, we align the beam of the subarray associated with $A$ with this tangent line towards $B$. Let $x=f(z)$ denote the desired trajectory of the bending beam, and $(x_B, z_B)$ denote the coordinate of point $B$ with $x_B = f(z_B)$. As such, the angle of departure (AoD) relative to the antenna boresight at point $A$ can be obtained as 
\begin{equation}
\theta_A = \arctan \frac{1}{f'(z_B)},
\end{equation}
or equivalently,
\begin{equation}
	\sin \theta_A = \frac{f'(z_B)}{\sqrt{1+\left(f'(z_B)\right)^{2}}}.
\end{equation}
Hence, the phase difference between $A$ and the endpoint of its subarray (i.e., $C$ in Fig. \ref{fig:Equivalent_rays}) should be $k\sin\theta_A{\text{d}}x$, where $k=\frac{2\pi}{\lambda}$. By this means, we can determine a continuous phase profile over the entire antenna array, denoted as $\phi(x)$, $0 \le x \le L$. Based on the above method, it should hold that \cite{curve_function}
\begin{equation}
	\label{equ:cal_phi}
	\frac{\text{d} \phi(x)}{\text{d} x}=k \frac{\text{d} f\left(z\right) / \text{d} z}{\sqrt{1+\left(\text{d} f\left(z\right) / \text{d} z\right)^{2}}}\text{.}
\end{equation}
 However, it is generally difficult to derive $\phi(x)$ in closed-form based on (\ref{equ:cal_phi}), as it is a complex differential equation for any given beam trajectory $x=f(z)$. To tackle this difficulty, the existing works always assume a sufficiently small AoD, such that $\sin\theta_A  \approx \tan\theta_A$ holds for any point $A$ within the antenna array, which leads to
\begin{equation}
	\label{equ:praxial}
	\frac{\text{d} f\left(z\right) / \text{d} z}{\sqrt{1+\left(\text{d} f\left(z\right) / \text{d} z\right)^{2}}} \approx \frac{\text{d} f\left(z\right)}{\text{d} z}.
\end{equation}
Note that for (\ref{equ:praxial}) to hold, the beam trajectory is required to have a sufficiently small curvature. Based on (\ref{equ:praxial}), (\ref{equ:cal_phi}) can be approximated as
\begin{equation}
	\label{equ:cla_phi_prax}
	\frac{\text{d} \phi(x)}{\text{d} x}=k \frac{\text{d} f\left(z\right)}{\text{d} z}.
\end{equation}
Consider a specific example of the parabolic-shaped beam, i.e., $x = \beta z^2$, where $\beta>0$ is a constant scalar. By substituting this function into (\ref{equ:cla_phi_prax}), we can obtain
\begin{equation}
	\label{equ:phase_shift}
	\phi(x)=-\frac{4}{3} \beta^{\frac{1}{2}} k x^{\frac{3}{2}}.
\end{equation}
\par It is worth noting that although the above method is capable of generating bending beams, it has several limitations. First, it assumes a continuous aperture with an infinite number of sub-arrays to achieve the desired beam trajectory, while the antennas can only be deployed at several discrete positions in practice, which may result in significant performance loss. Second, its phase-shift design only takes into account the direct link between each sub-array and its associated position along the trajectory (e.g., $A$ and $B$), while ignoring the interference from other sub-arrays to $B$. Third, its phase-shift design ignores the effects of the distance, as each position along the trajectory may experience varying path losses with the sub-arrays. Hence, it is desired that the beam trajectory is convex in $z$, such that more energy can be beamed to the far-away positions along the beam trajectory. Last but not least, this method only applies to \ac{ABF}, lacking in the amplitude design for \ac{DBF}. To overcome the above limitations, we propose a new and general bending beamforming method based on max-min beamforming, as detailed below.
\section{Proposed Bending Beamforming via Max-Min Beamforming}
\par In this section, we present the proposed bending beamforming method based on convex optimization techniques. In particular, we directly tackle the actual received power at each position along the trajectory by solving a max-min optimization problem.
 \subsection{Max-Min Beamforming}
\par Specifically, similar to Fig. \ref{fig:Equivalent_rays}, we assume that the antenna array is placed at the $x$-axis. Unlike the tangent method, we aim to maximize the minimum received signal power among all possible positions along the trajectory, so as to focus as much energy as possible on the trajectory. However, the beam trajectory is continuous and difficult to handle in the optimization. To circuvment this difficulty, we perform discrete sampling along the trajectory. Given the desired trajectory $x=f(z)$, let $M$ and $(x_m, z_m)$ denote the total number of sampling points and the coordinate of the $m$-th sampling point, with $x_m = f(z_m)$, $m \in {\cal M} \triangleq \{1,2,\cdots,M\}$.
\par As such, the channels between all antennas and the $M$ sampling points can be expressed by
 \begin{equation}
 	\bm{H} = [\bm{h}(x_1,z_1),\bm{h}(x_2,z_2),..,\bm{h}(x_M,z_M)]^\text{T}\in \mathbb{C}^{N\times M}\text{,}
 \end{equation}
 where we have omitted $y_m=0, m \in {\cal M}.$ for notational simplicity. The received signal powers at all of the $M$ sampling points can be expressed as
 \begin{equation}
 	\bm{p} = [p(x_1,z_1),p(x_2,z_2),...,p(x_M,z_M)]^\text{T}\in \mathbb{C}^{M\times 1},
 \end{equation}
with $p(x_m,z_m)=\lvert {\bm{\omega}}^\text{H}{\bm{h}}(x_m,z_m) \rvert^2, m \in {\cal M}$. Based on the above, the minimum received signal power among all sampling points can be expressed as
 \begin{equation}
 	p_{\text{min}} = \mathop {\min}\limits_{m \in \cal M} p(x_m,z_m)\text{.}
 \end{equation}
  If we take the \ac{ABF} as an example, the associated max-min optimization problem for bending beamforming is formulated as
\begin{align}
 			\text{(P1)}\quad & \max_{\bm{\omega}} \ p_{\text{min}} \nonumber \\
 			\text{s.t.} \quad &|\omega_a(n)| =  \frac{1}{\sqrt{N}}, \quad n \in {\cal N}\triangleq \left\{1, 2, \dots, N\right\}. \tag{18{a}} \label{Problem18a}
\end{align}
 Similarly, by changing \eqref{Problem18a} into $\sum_{n=1}^{N} |\omega_d(n)|^2 = 1$, (P1) becomes the max-min problem for \ac{DBF}.
 \par However, (P1) is a non-convex optimization problem that is challenging to be optimally solved. Although the \ac{SDR} method has been widely applied to solve this problem, it may suffer undesired local optimality and performance loss due to the Gaussian randomization procedure. In this paper, we propose a penalty-based algorithm to solve (P1) without the need for Gaussian randomization, as detailed next.
 \vspace{-4pt}
 \subsection{Proposed Solution to (P1)}
 \par First, we introduce an auxiliary variable $t$ to reformulate (P1) into its epigraph form, which can be expressed by
 		\begin{align}
 			\text{(P2)} \quad & \underset{\bm{\omega},t}{\text{max}}\:t, \nonumber \\
 			\text{s.t.} \quad & p(x_m, z_m)\geq t, \quad \tag{19{a}}\forall m\in {\cal M}, \label{Problem19a}\\
 						\quad &\eqref{Problem18a} \nonumber.
 		\end{align}
 To deal with the non-convex constraint in (\ref{Problem19a}), we rewrite the received signal power at each sampling point as
 \begin{equation}\tag{20}
 	\begin{split}
 	p_m = |\bm{\omega}^\text{H}\bm{h}_m\bm{h}_m^\text{H}\bm{\omega}|=\text{Tr}(\bm{R}_m\bm{V}),
 	\end{split}
 \end{equation}
 where $\bm{R}_m=\bm{h}_m\bm{h}_m^\text{H}$ and $\bm{V} =\bm{\omega}\bm{\omega}^\text{H}$, with $\text{rank}(\bm{V}) = 1$. By applying the \ac{SDR} technique to (P2), the optimization problem for \ac{ABF} can be transformed into
		\begin{align}
			 \text{(P3)} \quad & \underset{\bm{V},t}{\text{max}}\:t \nonumber \\
			\text{s.t.} \quad &\text{Tr}(\bm{R}_m\bm{V})  \geq t, \quad \forall m\in {\cal M} \tag{21{a}} \label{Problem21a}\\
			&\bm{V}(n,n) = \frac{1}{N},\quad \forall n \in {\cal N} \tag{21{b}} \label{Problem21b},
		\end{align}
For \ac{DBF}, (\ref{Problem21b}) can be replaced with Tr$(\bm{V})=1$.
\par Note that (P3) is a standard \ac{SDP} problem without the rank-one constraint $\text{rank}(\bm{V})=1$. Hence, it can be optimally solved by invoking the interior-point algorithm \cite{interpoint}. However, the optimized $\bm{V}$ by the \ac{SDP} may not satisfy the rank-one constraint. To address this issue, we propose a penalty-based approach that moves the rank-one constraint to the objective function as a regularization term. By introducing a penalty parameter, we introduce a penalty term for deviation from the rank-one requirement in the objective function, thereby guiding the solution toward rank-one matrices. To this end, note that the rank-one constraint can be equivalently expressed as a function of the trace and the maximum singular value of $\bm{V}$, i.e.,
 	\begin{equation}\tag{22}
 		\text{rank}(\bm{V}) = 1  \Leftrightarrow  f (\bm{V}) \triangleq \text{Tr}(\bm{V})-\sigma(\bm{V})=0, 
 	\end{equation}
where $\sigma(\bm{V})$ represents the maximum singular value of $\bm{V}$. As such, we can transform the objective function of (P3) into the following with a penalty term, i.e.,
\begin{equation}\tag{23}
	\underset{t,\bm{V}}{\text{max}}\,t-\rho f(\bm{V}),
	\label{equ:max_fun}
\end{equation}
where $\rho \geq 0$ denotes a penalty parameter introduced to force the objective function $ f(\bm{V})$ to approach zero if the rank-one constraint $\text{Tr}(\bm{V}) - \sigma(\bm{V}) = 0 $ is violated. However, introducing this penalty term renders the resulting objective function non-convex, as the maximum singular value $ \sigma(\bm{V}) $ is a convex (rather than concave) function. Nonetheless, \eqref{equ:max_fun} turns out to be the difference between an affine function and a concave function, making the \ac{SCA} algorithm applicable to solve it, by iteratively solving a series of convex subproblems to converge to a locally optimal solution \cite{SCA}. In particular, for a given local point \( \bm{V}^{(i)} \) in the \( i \)-th iteration of the \ac{SCA} algorithm, we approximate \( f(\bm{V}) \) using its first-order Taylor expansion, expressed as
\begin{equation}\tag{24}
	\begin{split}
	f(\bm{V})\geq \hat{f}(\bm{V}|\bm{V}^{(i)})\triangleq\text{Tr}&(\bm{V})-\rho(\bm{V}^{(i)})+\\
	&\text{Re}\left\{{\text{Tr}(\bm{s} \bm{s}^\text{H})(\bm{V}-\bm{V}^{(i)})}\right\},
\end{split}
\end{equation}
where $\bm{s}$ represents the singular vector corresponding to the maximum singular value of $\bm{V}^{(i)}$. Hence, the beamforming optimization in the $i$-th \ac{SCA} iteration is given by
		\begin{align*}
			\text{(P3-$i$)} \quad &\underset{\bm{V},t}{\text{max}}\:t-\hat{f}(\bm{V}|\bm{V}^{(i)}) \\
			\text{s.t.} \quad &\eqref{Problem21a},\:    
						\eqref{Problem21b}.
		\end{align*}
		Note that (P3-$i$) is a convex optimization problem, which can be optimally solved using the interior-point algorithm. Let ${\bm{V}}^*$ denote the optimal solution to (P3-$i$). Then, we can update ${\bm{V}}^{(i+1)}={\bm{V}}^*$ and proceed to solve (P3-$(i+1)$) accordingly.
\begin{algorithm}
	
	\caption{\ac{SCA} algorithm to solve (P1)}
	\textbf{Input:} ${\bm{V}}^{(0)}$. \\
	\textbf{Output:} $\bm{\omega}$.
	\begin{algorithmic}[1]
		\State Initialize: $i\gets0$.
		\While{convergence is not reached}
		\State Obtain $\bm{V}^*$ by solving problem (P3-$i$).
		\State Update ${\bm{V}}^{(i+1)}\gets\bm{V}^*$.
		\State $i \gets i + 1$.
		\EndWhile
		\State Retrieve $\bm{\omega}$ based on (\ref{equ:omega})
		\State \textbf{return} $\bm{\omega}$.
	\end{algorithmic}
\end{algorithm}
\par Let ${\bm{V}}_o$ denote the optimized solution to (P3) by the \ac{SCA} algorithm upon its convergence. Next, to retrieve the rank-one beamforming solution, we perform \ac{SVD} on ${\bm{V}}_o$ as ${\bm{V}}_o={\bm{U}}^H{\bm{\Omega}}{\bm{S}}$, where $\bm{U}$, ${\bm{\Omega}}$ and $\bm{S}$ are the left eigenvector matrix, the diagonal matrix of the singular values and the right eigenvector matrix of $\bm{V}_o$, respectively. Then we can obtain the beamforming solution to (P3) as
\begin{equation}\tag{26}
	\label{equ:omega}
	\bm{\omega} = \frac{1}{\sqrt{N}}e^{(j\text{arg}(\bm{U}^{\text{H}} \bm{\Lambda}^{1/2}\bm{r} ))}
\end{equation}
where $\bm{r}$ is the left singular vector corresponding to the largest singular value of $\bm{V}$. The main procedures for solving (P1) are summarized in \textbf{Algorithm 1}. It can be shown that the overall computational complexity of \textbf{Algorithm 1} is $O\left(MN^3\right)$ \cite{beck2010sequential}.
 \section{Numerical Results}
\par In this section, we present numerical results to evaluate the performance of the proposed method for generating bending beams. Unless otherwise stated, the simulation parameters are set as follows. The carrier frequency is $f_c=300$ GHz. The number of antennas at the \ac{BS} is $N=400$ and the spacing between any two adjacent antennas is set to half-wavelength.  As such, the Rayleigh distance can be calculated as 80 m. The user's position is located at $(x,z)=(225\beta,15)$. The antenna array is assumed to be located in the $x$-axis, and the beam trajectory is parabolic-shaped and given by $x=\beta z^2, 0 \leq z \leq 15$. The number of sampling points along the bending beam trajectory is set to $M = 200$. For performance comparision, we show the performance achieved by the proposed scheme under \ac{ABF} and \ac{DBF}, as well as the traditional \ac{TM} in \cite{6G_bending_beam}.
\begin{figure*}[htbp]
	\centering
	% 第一行
	\begin{subfigure}[b]{0.3\textwidth}
		\centering
		\includegraphics[width=\textwidth]{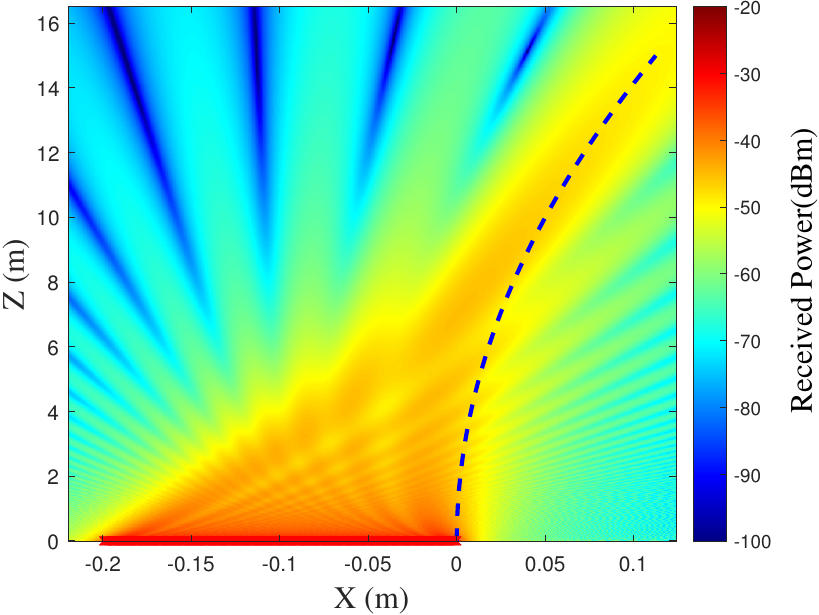}
		\caption{\ac{ABF}, $\beta =  0.0005$.}
		\label{fig:bending_beam2_a}
	\end{subfigure}
	\hspace{0.03\textwidth}
	\begin{subfigure}[b]{0.3\textwidth}
		\centering
		\includegraphics[width=\textwidth]{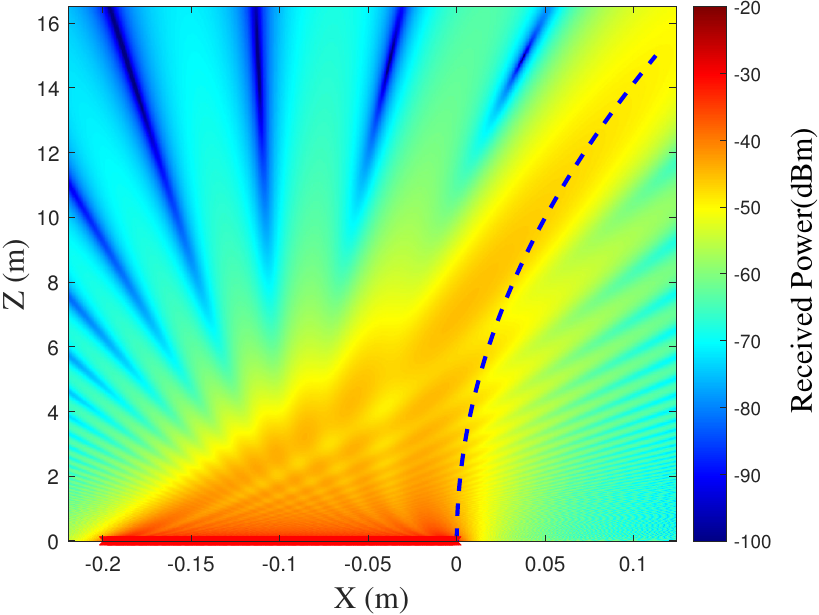}
		\caption{\ac{DBF}, $\beta =  0.0005$.}
		\label{fig:bending_beam2_d}
	\end{subfigure}
	\hspace{0.03\textwidth}
	\begin{subfigure}[b]{0.3\textwidth}
		\centering
		\includegraphics[width=\textwidth]{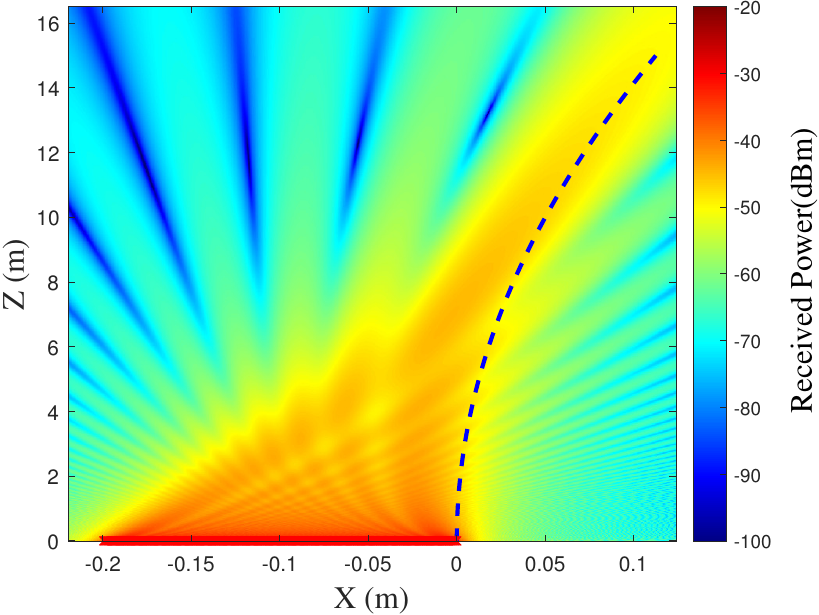}
		\caption{\ac{TM}, $\beta =  0.0005$.}
		\label{fig:bending_beam2}
	\end{subfigure}
	% 第二行
	\begin{subfigure}[b]{0.3\textwidth}
		\centering
		\includegraphics[width=\textwidth]{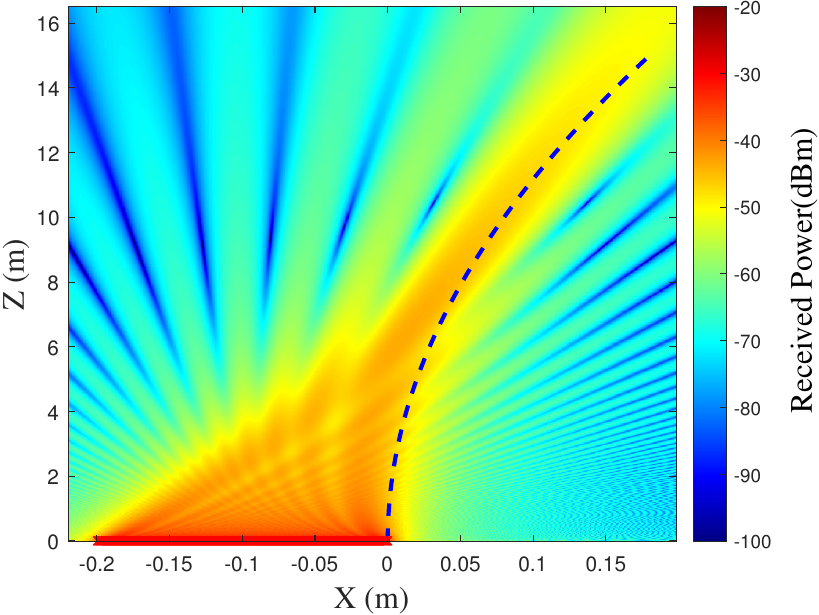}
		\caption{\ac{ABF}, $\beta = 0.0008$.}
		\label{fig:bending_beam4_a}
	\end{subfigure}
	\hspace{0.03\textwidth}
	\begin{subfigure}[b]{0.3\textwidth}
		\centering
		\includegraphics[width=\textwidth]{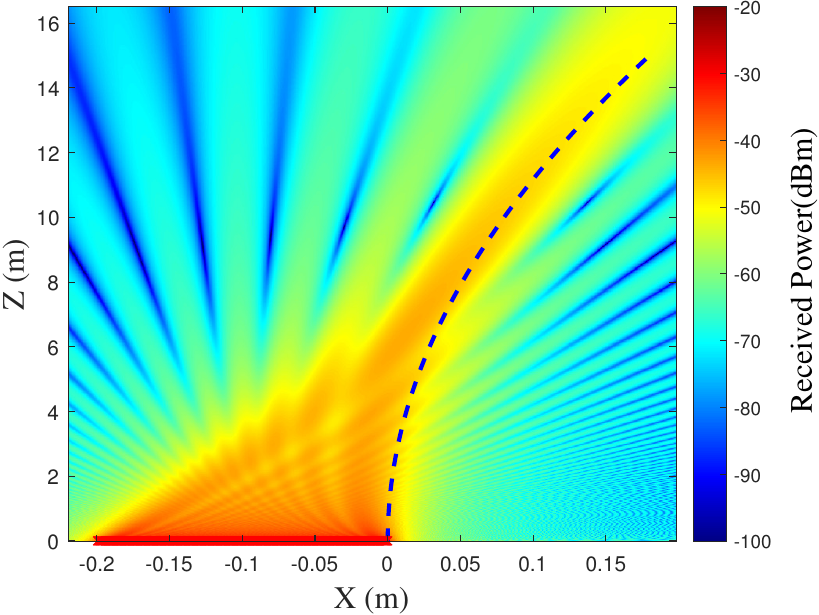}
		\caption{\ac{DBF}, $\beta = 0.0008$.}
		\label{fig:bending_beam4_d}
	\end{subfigure}
	\hspace{0.03\textwidth}
	\begin{subfigure}[b]{0.3\textwidth}
		\centering
		\includegraphics[width=\textwidth]{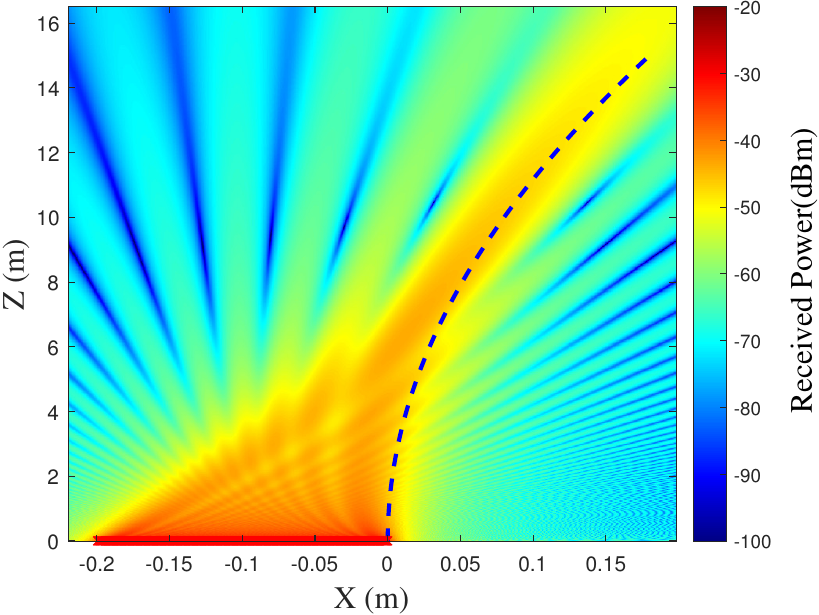}
		\caption{\ac{TM}, $\beta = 0.0008$.}
		\label{fig:bending_beam4}
	\end{subfigure}
	% 第三行
	\begin{subfigure}[b]{0.3\textwidth}
		\centering
		\includegraphics[width=\textwidth]{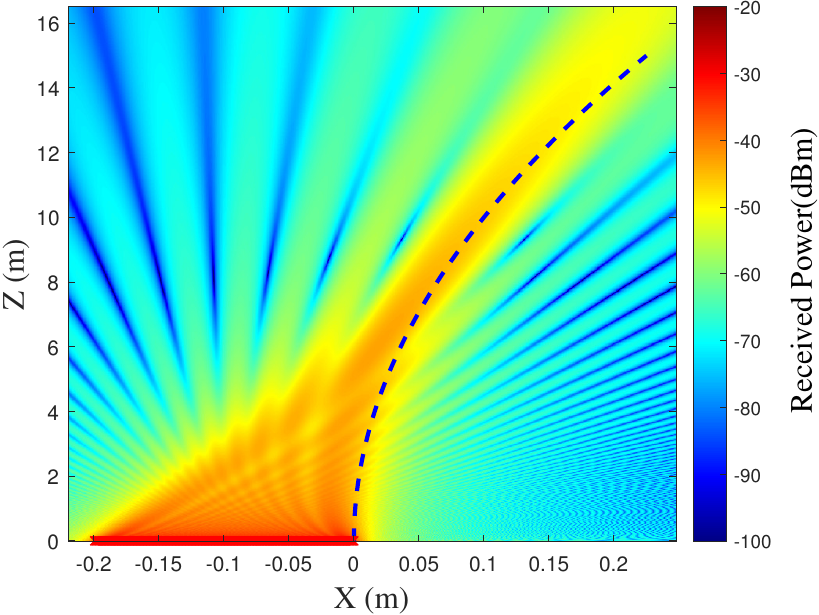}
		\caption{\ac{ABF}, $\beta = 0.001$.}
		\label{fig:bending_beam6_a}
	\end{subfigure}
	\hspace{0.03\textwidth}
	\begin{subfigure}[b]{0.3\textwidth}
		\centering
		\includegraphics[width=\textwidth]{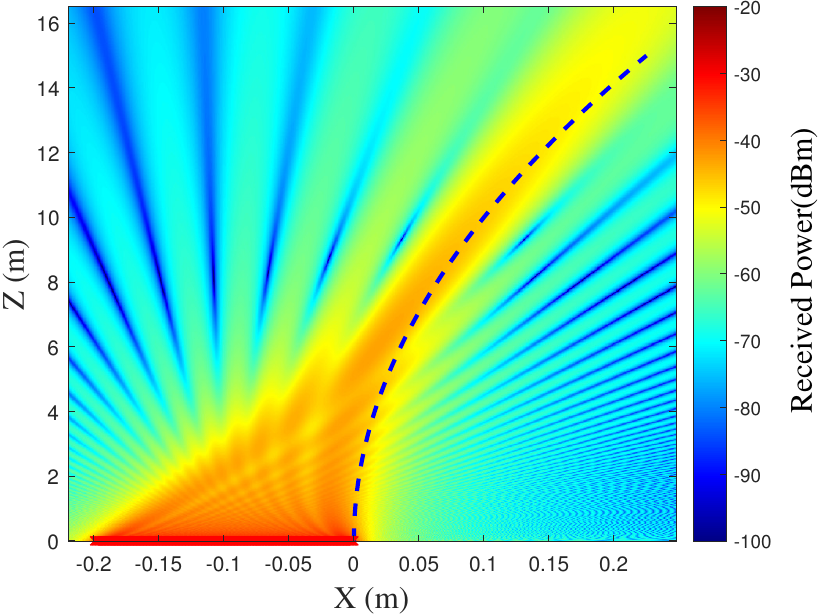}
		\caption{\ac{DBF}, $\beta = 0.001$.}
		\label{fig:bending_beam6_d}
	\end{subfigure}
	\hspace{0.03\textwidth}
	\begin{subfigure}[b]{0.3\textwidth}
		\centering
		\includegraphics[width=\textwidth]{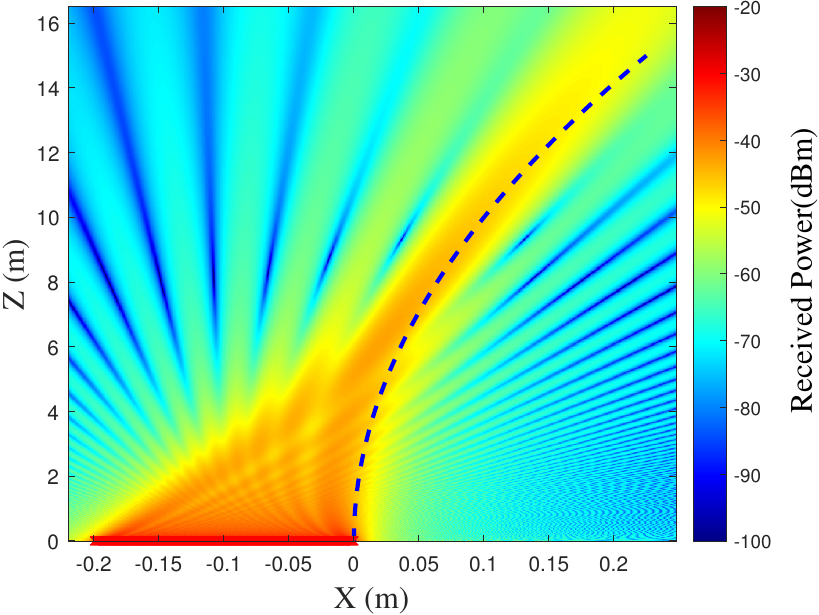}
		\caption{\ac{TM}, $\beta = 0.001$.}
		\label{fig:bending_beam6}
	\end{subfigure}
	\caption{Parabolic-shaped bending beams generated by different schemes for different values of $\beta$.}
	\label{fig:combined_bending_beams}
\end{figure*}
 \par First, Fig. \ref{fig:combined_bending_beams} plots the bending beams generated by different schemes for different values of $\beta$, where the desired beam trajectory is indicated by blue dashed lines. It is observed that for each $\beta$ considered, the proposed scheme yields high received signal power along the desired trajectory for both \ac{ABF} and \ac{DBF}. This result suggests that the beam generated by the proposed method exhibits non-diffracting characteristics similar to that by the conventional \ac{TM}, as shown in Fig. \ref{fig:bending_beam2}, Fig. \ref{fig:bending_beam4}, Fig. \ref{fig:bending_beam6}.
 \begin{figure*}[htbp]
 		\centering
 	\begin{subfigure}[b]{0.3\textwidth}
 		\centering
 		\includegraphics[width=\textwidth]{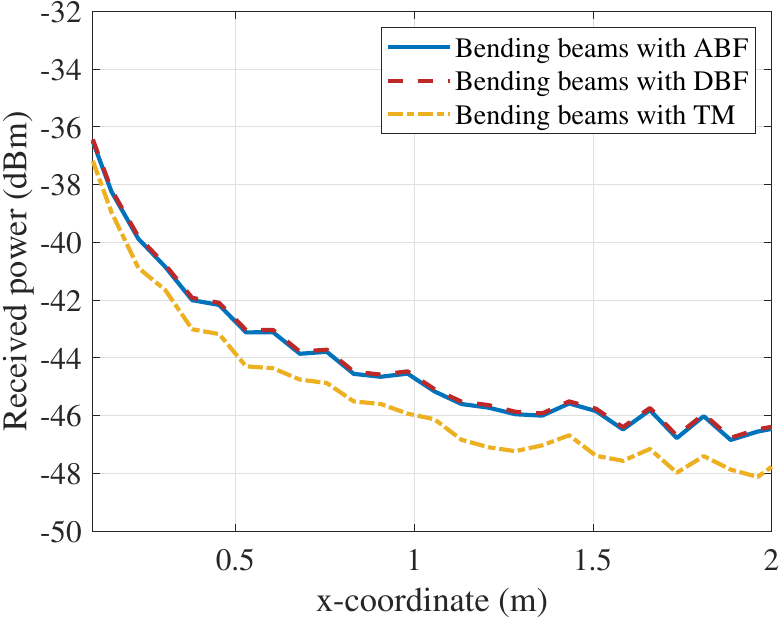} 
 		\caption{$\beta =  0.0005$.}
 		\label{fig:compare2} % 图片的标签，用于交叉引用
 	\end{subfigure}
 		\hspace{0.03\textwidth}
 	\begin{subfigure}[b]{0.3\textwidth}
 		\centering
 		\includegraphics[width=\textwidth]{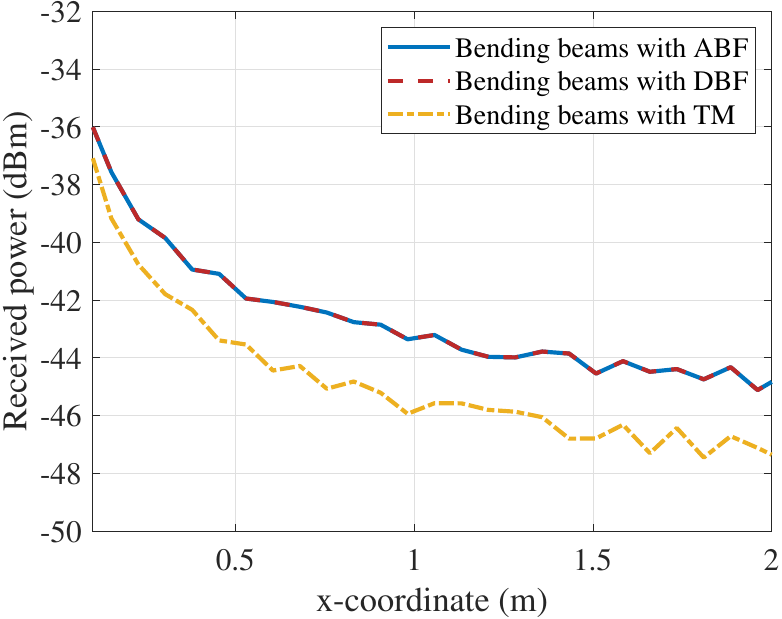} 
 		\caption{$\beta =  0.0008$.}
 		\label{fig:compare3} % 图片的标签，用于交叉引用
 	\end{subfigure}
 		\hspace{0.03\textwidth}
 	\begin{subfigure}[b]{0.3\textwidth}
 		\centering
 		\includegraphics[width=\textwidth]{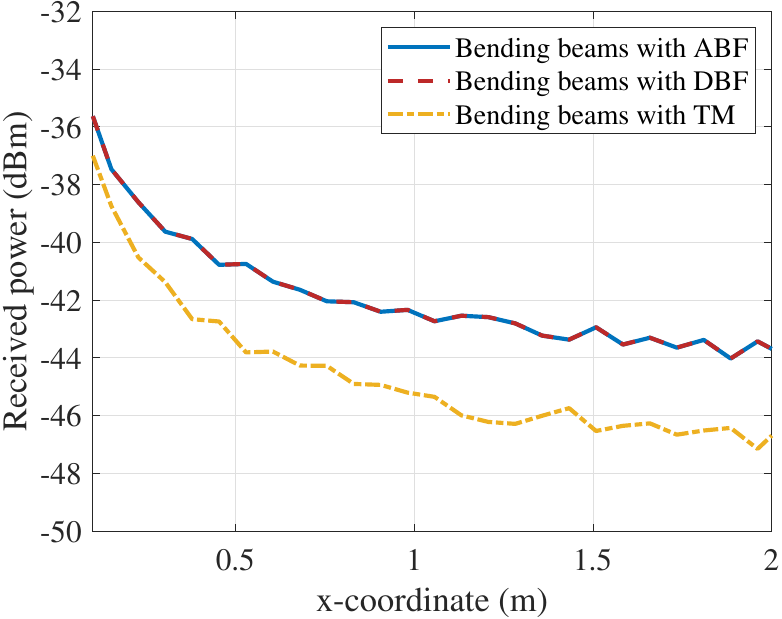} 
 		\caption{$\beta =  0.001$.}
 		\label{fig:compare6} % 图片的标签，用于交叉引用
 	\end{subfigure}
 	\caption{Distribution of the received signal power along the desired trajectory by different schemes.}
 	\label{fig:compare}
 \end{figure*}
 \par To provide a quantitative performance comparison between the proposed method and the \ac{TM}, we plot in Fig. \ref{fig:compare} the received signal power at all positions along the desired trajectory by different schemes. It is observed that compared to the \ac{TM}, the proposed max-min bending beamforming method achieves a more uniform distribution of the received signal power along the desired trajectory for both \ac{ABF} and \ac{DBF}, resulting in a higher max-min received signal power and received signal power at the user. It is also observed that the \ac{ABF} achieves a comparable performance to the \ac{DBF}. This observation aligns with the prior studies on wavefront engineering in \cite{6G_bending_beam}, which highlight the dominant role of phase control over amplitude control in generating bending beams. It is also observed that the performance gap between the proposed method and the \ac{TM} increases with $\beta$, suggesting that the proposed method is more preferred for trajectories with large curvature compared to the \ac{TM}. This is consistent with our discussion at the end of Section \ref{sec:tgb}.
 \par
\begin{figure}[!t]
	\centering
	\includegraphics[width=0.4\textwidth]{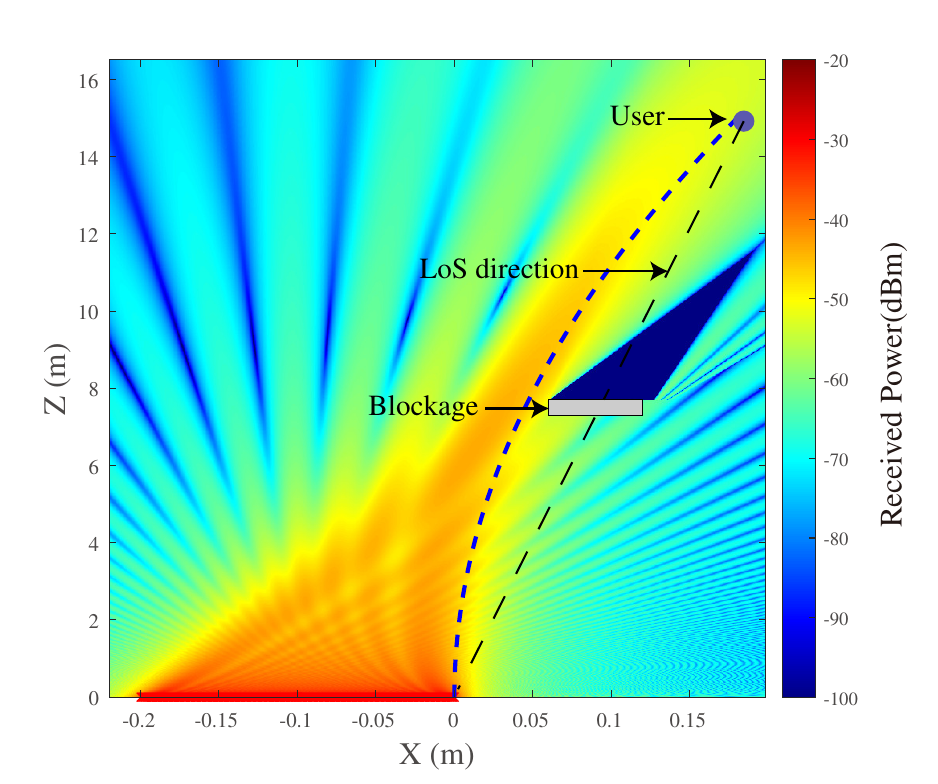} 
	\caption{Bending beam generated by the proposed method in the presence of a blockage.}
	\label{fig:blockage} % 图片的标签，用于交叉引用
	\end{figure}

 In the previous examples, we show the performance of the proposed method without accounting for actual obstacles in the environment. To evaluate its robustness in the presence of obstacles, we show in Fig. \ref{fig:blockage} the bending beam generated under \ac{ABF}, where an obstacle is placed between the antenna array and the user. If the \ac{LoS} channel between the $n$-th antenna and any position $(x,z)$ is blocked by the obstacle, we set $h_n(x,z)=0$. The desired beam trajectory is set to $x=0.0008z^2, 0 \le z \le 15 $m. As observed from Fig. \ref{fig:blockage}, even the obstacle can nullify the \ac{LoS} channels between several antennas and positions, the generated bending beam is able to maintain a pre-defined shape without significant deviation, showing high robustness against obstructions. It thus follows that the proposed max-min bending beamforming method offers an efficient solution to resolve the challenging issue of \ac{THz} signal blockage.
 
  \section{Conclusion}
\par In this paper, we proposed a new and general bending beamforming method by formulating a max-min beamforming approach. Unlike the conventional tangent method, our proposed method directly tackles the actual received signal power and significantly reduces the reliance the geometric properties of the desired trajectory. To solve the non-convex max-min optimization problem in the proposed method, we adopted the \ac{SCA} algorithm to obtain a locally optimal solution. Simulation results demonstrated that the proposed method achieves superior performance over the tangent method in terms of the max-min received signal power and the received signal power at the user.
\bibliographystyle{IEEEtran}
\bibliography{reference}

\begin{thebibliography}{10}
\providecommand{\url}[1]{#1}
\csname url@samestyle\endcsname
\providecommand{\newblock}{\relax}
\providecommand{\bibinfo}[2]{#2}
\providecommand{\BIBentrySTDinterwordspacing}{\spaceskip=0pt\relax}
\providecommand{\BIBentryALTinterwordstretchfactor}{4}
\providecommand{\BIBentryALTinterwordspacing}{\spaceskip=\fontdimen2\font plus
\BIBentryALTinterwordstretchfactor\fontdimen3\font minus
  \fontdimen4\font\relax}
\providecommand{\BIBforeignlanguage}[2]{{%
\expandafter\ifx\csname l@#1\endcsname\relax
\typeout{** WARNING: IEEEtran.bst: No hyphenation pattern has been}%
\typeout{** loaded for the language `#1'. Using the pattern for}%
\typeout{** the default language instead.}%
\else
\language=\csname l@#1\endcsname
\fi
#2}}
\providecommand{\BIBdecl}{\relax}
\BIBdecl

\bibitem{boyu_thz}
B.~Ning, Z.~Tian, W.~Mei, Z.~Chen, C.~Han, S.~Li, J.~Yuan, and R.~Zhang,
  ``Beamforming technologies for ultra-massive {MIMO} in {T}erahertz
  communications,'' \emph{IEEE Open J. Commun. Soc.}, vol.~4, pp. 614--658,
  2023.

\bibitem{pencil_like_beam}
J.~Ye, H.~Gharavi, and B.~Hu, ``Fast beam discovery and adaptive transmission
  under frequency selective attenuations in sub-{T}erahertz bands,'' \emph{IEEE
  Trans. Signal Process.}, vol.~71, pp. 727--740, 2023.

\bibitem{300GHz_100m}
Z.~Chen, X.~Ma, B.~Zhang, Y.~Zhang, Z.~Niu, N.~Kuang, W.~Chen, L.~Li, and
  S.~Li, ``A survey on {T}erahertz communications,'' \emph{China
  Communications}, vol.~16, no.~2, pp. 1--35, 2019.

\bibitem{Ultra-Dense_Network1}
M.~Kamel, W.~Hamouda, and A.~Youssef, ``Ultra-dense networks: A survey,''
  \emph{IEEE Commun. Surveys Tuts.}, vol.~18, no.~4, pp. 2522--2545, 2016.

\bibitem{Ultra-Dense_Network2}
Y.~Teng, M.~Liu, F.~R. Yu, V.~C.~M. Leung, M.~Song, and Y.~Zhang, ``Resource
  allocation for ultra-dense networks: A survey, some research issues and
  challenges,'' \emph{IEEE Commun. Surveys Tuts.}, vol.~21, no.~3, pp.
  2134--2168, 2019.

\bibitem{Intelligent_Reflecting_Surface}
W.~Mei, B.~Zheng, C.~You, and R.~Zhang, ``Intelligent reflecting surface-aided
  wireless networks: From single-reflection to multireflection design and
  optimization,'' \emph{Proc. IEEE.}, vol. 110, no.~9, pp. 1380--1400, 2022.

\bibitem{weidong_ris}
B.~Zheng, C.~You, W.~Mei, and R.~Zhang, ``A survey on channel estimation and
  practical passive beamforming design for intelligent reflecting surface aided
  wireless communications,'' \emph{IEEE Commun. Surveys Tuts.}, vol.~24, no.~2,
  pp. 1035--1071, 2022.

\bibitem{Helmholtz_equ}
G.~A. Siviloglou, J.~Broky, A.~Dogariu, and D.~N. Christodoulides,
  ``Observation of accelerating {A}iry beams,'' \emph{Phys. Rev. Lett.},
  vol.~99, p. 213901, Nov 2007.

\bibitem{wavefont}
A.~Singh, V.~Petrov, H.~Guerboukha, I.~V. Reddy, E.~W. Knightly, D.~M.
  Mittleman, and J.~M. Jornet, ``Wavefront engineering: Realizing efficient
  {T}erahertz band communications in 6{G} and beyond,'' \emph{IEEE Wireless
  Commun.}, vol.~31, no.~3, pp. 133--139, 2024.

\bibitem{wavefonthopping}
V.~Petrov, H.~Guerboukha, D.~M. Mittleman, and A.~Singh, ``Wavefront hopping:
  An enabler for reliable and secure near field terahertz communications in
  6{G} and beyond,'' \emph{IEEE Wireless Commun.}, vol.~31, no.~1, pp. 48--55,
  2024.

\bibitem{nature_comunication}
H.~Guerboukha, B.~Zhao, Z.~Fang, E.~Knightly, and D.~M. Mittleman, ``Curving
  {THz} wireless data links around obstacles,'' \emph{Commun. Eng.}, vol.~3,
  no.~1, p.~58, Mar 2024.

\bibitem{6G_bending_beam}
S.~Droulias, G.~Stratidakis, and A.~Alexiou, ``Bending beams for 6{G}
  near-field communications,'' \emph{IEEE Trans. Wireless Commun.}, vol.~24,
  no.~2, pp. 1467--1480, 2025.

\bibitem{curve_function}
L.~Froehly, F.~Courvoisier, A.~Mathis, M.~Jacquot, L.~Furfaro, R.~Giust,
  P.~Lacourt, and J.~Dudley, ``Arbitrary accelerating micron-scale caustic
  beams in two and three dimensions,'' \emph{Opt. Express.}, vol.~19, no.~17,
  pp. 16\,455--16\,465, 2011.

\bibitem{interpoint}
B.~Ning, W.~Mei, L.~Zhu, Z.~Chen, and R.~Zhang, ``Codebook design and
  performance analysis for wideband beamforming in terahertz communications,''
  \emph{IEEE Trans. Wireless Commun.}, vol.~23, no.~12, pp. 19\,618--19\,633,
  2024.

\bibitem{SCA}
S.~Boyd and L.~Vandenberghe, \emph{Convex optimization}.\hskip 1em plus 0.5em
  minus 0.4em\relax Cambridge, U.K.: Cambridge Univ. Press, 2009.

\bibitem{beck2010sequential}
A.~Beck, A.~Ben-Tal, and L.~Tetruashvili, ``A sequential parametric convex
  approximation method with applications to nonconvex truss topology design
  problems,'' \emph{Journal of Global Optimization}, vol.~47, pp. 29--51, 2010.

\end{thebibliography}
\end{document}